\documentclass[letterpaper,12pt]{article}

\usepackage{mathptmx}
\usepackage{amssymb,amsmath}
\usepackage{hyperref}
\usepackage{color,soul}

\usepackage{graphics}
\usepackage{graphicx}

\usepackage[authoryear,comma,longnamesfirst,sectionbib]{natbib} 

\usepackage{background}

\definecolor{textcolor}{HTML}{0A75A8}
\newcommand\Text{preprint, to appear in Statistical Applications in Genetics and Molecular Biology}

\SetBgColor{textcolor}
\SetBgOpacity{1}
\SetBgAngle{90}
\SetBgPosition{-0.6,-10}
\SetBgScale{1}
\SetBgContents{\sffamily\Text}

\begin{document}


\title{Genome-wide association studies with high-dimensional phenotypes}

\author{Pekka Marttinen$^1$, Jussi Gillberg$^1$, Aki Havulinna$^2$, \\
Jukka Corander$^3$, and Samuel Kaski$^{1,4}$}


\date{}

\maketitle

\begin{center}
{\small $^1$Helsinki Institute for Information Technology HIIT, Department of Information and Computer Science, Aalto University, Espoo, Finland, $^2$Department of Chronic Disease Prevention, National Institute for Health and Welfare, Helsinki, Finland, $^3$Helsinki Institute for Information Technology HIIT, Department of Mathematics and Statistics, University of Helsinki, Helsinki, Finland, $^4$Helsinki Institute for Information Technology HIIT, Department of Computer Science, University of Helsinki, Helsinki, Finland}
\end{center}

\pagebreak

\begin{abstract}
High-dimensional phenotypes hold promise for richer findings in association studies, but testing of several phenotype traits aggravates the grand challenge of association studies, that of multiple testing. Several methods have recently been proposed for testing jointly all traits in a high-dimensional vector of phenotypes, with prospect of increased power to detect small effects that would be missed if tested individually. However, the methods have rarely been compared to the extent of enabling assessment of their relative merits and setting up guidelines on which method to use, and how to use it. We compare the methods on simulated data and with a real metabolomics data set comprising 137 highly correlated variables and approximately 550,000 SNPs. 

Applying the methods to genome-wide data with hundreds of thousands of markers inevitably requires division of the problem into manageable parts facilitating parallel processing, parts corresponding to individual genetic variants, pathways, or genes, for example. Here we utilize a straightforward formulation according to which the genome is divided into blocks of nearby correlated genetic markers, tested jointly for association with the phenotypes. This formulation is computationally feasible, reduces the number of tests, and lets the methods take advantage of combining information over several correlated variables not only on the phenotype side, but also on the genotype side.

Our experiments show that canonical correlation analysis has higher power than alternative methods, while remaining computationally tractable for routine use in the GWAS setting, provided the number of samples is sufficient compared to the numbers of phenotype and genotype variables tested. Sparse canonical correlation analysis and regression models with latent confounding factors show promising performance when the number of samples is small compared to the dimensionality of the data.
\end{abstract}

\pagebreak

\section{Introduction}

The increasingly widely collected 'omics' data, including genomics, transcriptomics, metabolomics, and proteomics data sets, bring completely new opportunities to genome-wide association studies (GWASs). GWAS searches for associations between the genome, typically represented as the single nucleotide polymorphisms (SNPs), and one phenotype variable, or trait. Examples of traits could be a disease indicator (binary, dichotomous) or the height of an individual (continuous-valued). The statistical challenge of the GWASs is the required multiple testing correction to account for the large number of associations ($>\!10^6$) to be tested, see, e.g., Balding (2006)\nocite{balding2006tutorial}. There is now a growing interest in detecting associations between genomics and the other types of omics data, where the phenotype is multivariate in contrast to the classical setting. For example, associations between genotypes and gene expressions (transcriptomics) have been studied, see, e.g., Parts et al. (2011) and Fusi et al. (2012); \nocite{parts2011joint,fusi2012joint} more recently, the genotypes have been associated with metabolomics phenotypes (Tukiainen et al., 2012; Inouye et al., 2012\nocite{tukiainen2012detailed,inouye2012novel}). With such studies, the problems related to the multiple testing become accentuated due to the growing dimensionality of the phenotype vector.

To tackle the statistical challenge in the omics-omics type association studies, several alternatives have been proposed. The simplest and most commonly used approach is to test associations between each genotype-phenotype pair in turn, and then to apply a stringent significance cutoff to account for the vast number of tests performed. For high-dimensional phenotypes this approach requires a very large sample size, however, as it fails to utilize the fact that a SNP affecting a phenotype variable is likely to have an effect also on other phenotypic variables which are highly correlated with the first one. The most straightforward modification to the simple pairwise testing is a combination of several pairwise tests, at the simplest by taking an average, for example.

However, it has been argued by Kim and Xing (2009a)\nocite{kim2009statistical} that accounting for the correlations between multiple phenotypes while testing for association is preferable to combining results from several related experiments in an \textit{ad hoc} manner after the pairwise testing. Kim and Xing (2009a,b)\nocite{kim2009statistical, kim2009tree} used sparse regression models for multiple correlated traits. The models favor sparsity in the regression coefficients while encouraging sharing of common regressors for correlated traits. Another class of statistical models that has been proposed for GWASs of high-dimensional phenotypes exploits latent variables to account for hidden confounders that, if unaccounted, would blur the analysis by causing false positive findings and reduced power. These latent variable regression models have earlier been successfully applied in association studies of gene expression measurements by Stegle et al. (2010) and Fusi et al. (2012)\nocite{stegle2010bayesian,fusi2012joint}. Canonical correlation analysis (CCA) is yet another multivariate technique that 
has recently been suggested as a tool for analyzing high-dimensional phenotypes in genome-wide association studies (Ferreira and Purcell, 2009)\nocite{ferreira2009multivariate}. CCA is a generalization of multivariate regression, where, instead of testing for an association between a pair of variables, an association between two groups of variables is tested (Hotelling, 1936\nocite{hotelling1936relations}). CCA is conceptually related to the aforementioned latent variable regression models, as it can be cast into a probabilistic formulation according to which the so-called canonical correlation between the two groups of variables is explained by hidden factors affecting simultaneously both sets of variables (Bach and Jordan, 2005\nocite{bach2005probabilistic}). Sparse CCA has been proposed for situations in which the dimension of a data set is too large to fit the classical CCA (Parkhomenko et al., 2009; Witten and Tibshirani, 2009).\nocite{parkhomenko2009sparse,witten2009extensions}

Given the list of methods, a researcher planning to carry out a GWAS with a high-dimensional phenotype immediately faces the challenge of choosing the method most appropriate for the data. The challenge is particularly hard as the methods have been developed and presented independently, and partly even without acknowledging each other. Thus, definite guidelines, let alone an established strategy on how to perform a statistical analysis in a GWAS with a high-dimensional phenotype are clearly lacking. Our goal in this paper is to investigate the suitability of various alternatives for GWAS with high-dimensional phenotypes. As the case study we use a recently published metabolomics data set of 137 quantitative traits with a genome-wide genotype data comprising approximately half a milloin single nucleotide polymorphisms sampled from 509 unrelated individuals (Inouye et al., 2010\nocite{inouye2010metabonomic}). These data are particularly suitable for our purpose because the ground truth is available as the data have earlier been 
analysed for associations as a part of a larger data set (Tukiainen et al., 2012; Inouye et al., 2012\nocite{tukiainen2012detailed,inouye2012novel}).

The genome is inherited as continuous chunks of DNA, resulting in high correlations (linkage disequilibrium) between neighboring SNPs, see, e.g., Frazer et al. (2007)\nocite{frazer2007second}. To make the methods computationally feasible in practice, instead of analyzing all SNPs together, we reduce the dimensionality by exploiting this characteristic of genotype data by dividing the genome into blocks of correlated SNPs (referred to as LD-blocks in the remainder of this paper) which we analyse separately (and in parallel) from each other. While joint analysis of the whole-genome genotype vector could be more accurate in principle, this blockwise approach is favourable in two respects: first, the number of tests is reduced, and second, the methods will be able to borrow information not only over correlated phenotypes, but also over several neighboring SNPs (for example if the causal SNP happens not to be present in the data set). The LD-block structure has previously been utilized in association studies of univariate traits (see discussion by Balding, 2006) and also to impute missing genotypes (Marchini and Howie, 2010)\nocite{marchini2010genotype}.

To summarize, we compare a set of recently developed sophisticated methods for realistic-sized GWAS with high-dimensional phenotypes. Based on experiences from the analysis of the real data along with comprehensive simulations, we provide practical suggestions on how data sets with similar characteristics can most appropriately be analyzed.

\section{Methods}

The main focus of this paper is to investigate how well different methods are able to take advantage of a joint analysis of all phenotypes in an association study. Here, we briefly outline the methods that we include in our study. We utilize the LD-block structure to enable the methods to account for information over several correlated SNPs. Thus, the association test scores provided by different methods are for LD-blocks, not individual SNPs, the idea being first to detect associated regions, followed by investigation of SNP-wise weights provided by the methods.\footnote{We note the argument of Donnelly (2008\nocite{donnelly2008progress}) that even fine mapping of candidate regions is unlikely to point to just one potentially causal SNP and, instead, will typically narrow researchers' attention to a set of such SNPs to be studied further in functional assays.} Consequently, if only SNP-wise scores are available from some method, these are combined into a single score in a straightforward manner, for example by taking the maximum or average over the block, as explained in more detail below. We apply multiple-testing correction on the block level by considering the maximum score over all blocks in permuted data sets.

\subsection{Canonical correlation analysis}

Canonical correlation analysis (CCA) is a multivariate technique designed for detecting associations between two groups of variables (Hotelling, 1936\nocite{hotelling1936relations}; see also Mardia, 1979; Hardoon et al., 2004\nocite{mardia1979multivariate,hardoon2004canonical}). Letting $X$ and $Y$ denote the $n \times q$ and $n \times p$ genotype and phenotype matrices and assuming without loss of generality that they are centered, the goal of CCA is to find a linear combination of the columns of $Y$ and a linear combination of the columns of $X$ that are maximally correlated with each other. This corresponds to finding vectors $a\in\mathbf{R}^q$ and $b\in\mathbf{R}^p$ such that
\begin{equation}\label{eq:can_cor}
\rho(a,b)=\frac{(Xa)\cdot(Yb)}{\|Xa\|\,\|Yb\|}
\end{equation}
becomes maximized. In CCA terminology, $Xa$ is called the best linear predictor and $Yb$ the most predictable criterion, although the underlying mathematics is symmetric.

Denoting by $S_x$ and $S_y$ the sample covariance matrices and $S_{xy}$ the cross-covariance matrix, the procedure for finding $a$ and $b$ starts by computing
\begin{equation}
K=S_{x}^{-1/2}S_{xy}S_{y}^{-1/2}
\end{equation}
and
\begin{equation}
N_1=KK',\qquad N_2=K'K.
\end{equation}
Then, by the singular value decomposition theorem, $K$ can be written as
\begin{equation}
K=(\alpha_1,\ldots,\alpha_k)D(\beta_1,\ldots,\beta_k)',
\end{equation}
where $\alpha_i$ and $\beta_i$ are the standardized eigenvectors of $N_1$ and $N_2$, $D=\text{diag}(\lambda_1^{1/2},\ldots,\lambda_k^{1/2})$ is a diagonal matrix of non-zero eigenvalues of $N_1$ (or $N_2$), and $k=\text{rank}(K)$. Now,
\begin{equation}
a_i=S_x^{-1/2}\alpha_i, \quad\text{and}\quad b_i=S_y^{-1/2}\beta_i
\end{equation}
are termed the $i$th canonical correlation vectors for X and Y, respectively. The objective (\ref{eq:can_cor}) is maximized by selecting $a=a_1$ and $b=b_1$. Furthermore, $a_2$ and $b_2$ are the coefficient vectors that maximize the correlation under the constraint that the resulting linear combinations are uncorrelated with $Xa$ and $Yb$, and so forth.

Canonical correlation analysis has recently been investigated in the context of genome-wide association studies (Ferreira and Purcell, 2009; Naylor et al., 2010; Tang and Ferreira, 2012\nocite{ferreira2009multivariate, naylor2010using, tang2012gene}). The difference between these articles is in how the problem is formulated in order to apply the CCA. Basically, one is left with the freedom of choosing the groups of variables between which associations are investigated. Naylor et al. (2010)\nocite{naylor2010using} divided their gene expression measurements into groups of three consecutive probes which were tested for association with SNPs located in the corresponding genomic region. Tang and Ferreira (2012)\nocite{tang2012gene} used genes with pruning to remove highly correlated SNPs to define the groups of SNPs to be tested for association with a relatively low-dimensional phenotype ($\leq\!6$). Recently, Inouye et al. (2012)\nocite{inouye2012novel} tested each SNP individually with groups of phenotypes, where the phenotype groups corresponded to clusters of highly correlated metabolites. In this paper, we consider two alternative strategies for defining the 
SNP groups: (1) selecting blocks of neighboring SNPs which are highly correlated due to linkage disequilibrium (the method being referred to as \textit{CCA-block} henceforth), or (2) analysing each SNP in the block individually and taking the maximum of the individual scores as the score of the block (\textit{CCA-single}). As the phenotype, we use all the metabolites jointly; in the simulations we additionally consider analysing only a subset of highly correlated metabolites at a time.

Two different approximations for determining the statistical significance of whether any of the $k$ correlations $\rho_i$ is non-zero have been utilized in genetic association studies. Both are based on the likelihood ratio statistic $\lambda=\prod_{i=1}^{k}(1-\rho_i^2)$, which has a Wilks $\Lambda(p, n-1-q, q)$ distribution (see, e.g., Mardia, 1979). \nocite{mardia1979multivariate} The first one uses Bartlett's approximation (see, e.g., Bartlett, 1941\nocite{bartlett1941statistical})
\begin{equation}
-(n-1-(p+q+1)/2)\ln \prod_{i=1}^k (1-\rho_i^2) \sim \chi^2_{pq}.
\end{equation} The second one uses Rao's F-approximation:
\begin{equation}
F_{(df_1,df_2)} = \left( \frac{1-\lambda^{1/s}}{\lambda^{1/s}} \right)\times\left(\frac{df_2}{df_1}\right),
\end{equation}
where
\begin{equation}
s=\sqrt{ \frac{p^2q^2-4}{p^2+q^2-5} }, \quad df_1=pq,
\end{equation}
and 
\begin{equation}
df_2=\left(\frac{2n-3-p-q}{2}\right)s-\frac{pq}{2}+1.
\end{equation}
Unless stated otherwise, the score for \textit{CCA-block} is taken to be the value of the maximum canonical correlation. Alternatively one could use (minus logarithm of) the p-value calculated for the block using either Bartlett's or Rao's approximations. However, we present some comparisons of these three possible approaches. Note that with \textit{CCA-single} only one canonical correlation may be calculated, giving a one-to-one mapping between the (maximum) canonical correlation and the corresponding p-values.

\subsection{Sparse canonical correlation analysis}
The classical CCA presented in the previous section assumes that the number of samples $n$ is larger than the dimension of either of the data sets $q$ or $p$. In practice, if $q$ or $p$ is close to $n$, one can always find a linear combination having canonical correlation near unity. To overcome the problem, several variants have been presented (Waaijenborg et al., 2008; Parkhomenko et al., 2009; Witten and Tibshirani, 2009; Chen et al., 2012). Common to these approaches is that they favor vectors of linear combination coefficients $a$ and $b$ in which some of the elements are equal to zero, for example by assuming a Lasso-type penalty for $a$ and $b$ (Tibshirani, 1996). As a representative sparse CCA method we use the implementation from Parkhomenko et al. (2009), although many of the methods appear to be closely related (see a detailed discussion in Witten and Tibshirani, 2009, for example). 

The idea of Parkhomenko (2009) is to introduce soft-thresholding parameters $\lambda_a$ and $\lambda_b$ for variable selection from sets $X$ and $Y$, respectively. The parameters control how many variables (columns) of each type will be included in the solution. During the iterative algorithm used for estimating $a$ and $b$, the coefficients whose absolute value is below the limit specified by the thresholding parameters are set to zero. The values of the thresholding parameters are determined using $k$-fold cross-validation over a grid of ($\lambda_a$,$\lambda_b$) values. The values that maximize the canonical correlation in the test sample are selected. In our experiments, we run the algorithm with default options for the grid values and $k$.

Since sparse CCA has been introduced in the genetic context it has been used to compute the canonical correlation between all phenotypes and either all SNPs or all SNPs in a single chromosome at a time. However, such strategies become computationally infeasible as the numbers of SNPs and samples increase. We investigate two approaches of using sparse CCA for scoring LD-blocks. The first one is obtained by simply calculating the canonical correlation between all SNPs in a block and all phenotypes using sparse CCA. However, it turns out that this strategy does not work very well (see \textit{Results}). As an alternative, we consider an approach where the genome is divided into larger windows by concatenating neighboring LD-blocks such that the number of SNPs within a window exceeds a given threshold (we used 2,000 SNPs in our experiments). Sparse CCA is then applied to the window of SNPs and all phenotypes. After this, the score for each LD-block within the window is obtained by calculating the classical canonical correlation value between all SNPs within the block having non-zero coefficients, and all phenotypes. The difference between this second strategy and \textit{CCA-block} is that sparsity is enforced among the SNPs in the block by utilizing only the SNPs with non-zero coefficients from the initial window-wide sparse CCA.

\nocite{waaijenborg2008quantifying, witten2009extensions, parkhomenko2009sparse, chen2012structured, tibshirani1996regression}

\subsection{Sparse regression for multiple correlated traits}

In the machine learning community, several regularized regression methods have recently been introduced for modeling correlated phenotypes, see, for instance, Kim and Xing (2009a,b) and Sohn and Kim (2012)\nocite{kim2009statistical, kim2009tree, sohn2012joint}. In these methods, the columns $\mathbf{y_k}$ of the phenotype matrix $Y$ are often modeled using separate regression models 
\begin{equation}
\mathbf{y_k} = X\mathbf{\beta_k} + \mathbf{\epsilon_k}, \quad k=1,\ldots,p.
\end{equation}
One popular method, GFlasso (Kim and Xing, 2009a\nocite{kim2009statistical}), facilitates borrowing of information between correlated phenotypes through learning the parameters with penalized least squares\footnote{This particular formulation is termed G$_w^2$Flasso by \cite{kim2009statistical}. We used cutoff 0.7 in our analyses to define the edges $E$ in the correlation graph.}:
\begin{equation}
\label{eq:gflasso_objective}
\begin{split}
\hat{B}& = \text{argmin} \sum_k (\mathbf{y_k}-X\mathbf{\beta_k})^T (\mathbf{y_k}-X\mathbf{\beta_k}) + \\
& \lambda \sum_j \sum_k |\beta_{jk}| + \gamma\sum_{(m,l)\in E}r_{ml}^2\sum_j|\beta_{jm}-\text{sign}(r_{ml})\beta_{jl}|,
\end{split}
\end{equation}
where $\hat{B}$ contains jointly the estimated parameter vectors $\mathbf{\beta_k}$. In (\ref{eq:gflasso_objective}), there are two regularization parameters, $\lambda$ and $\gamma$, to be learned through cross-validation. The purpose of the parameter $\lambda$ is to shrink the coefficients towards zero, favoring models with few non-zero coefficients. The term including the $\gamma$ parameter has been included to favor sharing information; it encourages the sizes of the effects $\beta_{jm}$ and $\beta_{jl}$ of SNP $j$ on correlated phenotypes $m$ and $l$ to be similar. In Equation (\ref{eq:gflasso_objective}), $r_{ml}$ is the correlation between the $m$th and $l$th phenotypes and $E$ is an \textit{a priori} specified phenotype correlation graph with edges representing correlations to be accounted for in the model.

\subsection{Regression with hidden confounding factors}

High-dimensional phenotypes are often correlated due to hidden confounders not related to genetic factors. For example, gene expression measurements may be affected by environmental conditions and experimental procedures (Leek and Storey, 2007; Gibson, 2008\nocite{leek2007capturing,gibson2008environmental}), which, unaccounted, would cause reduced power and increased false positive rate in association studies. To handle such confounders, several methods where the hidden determinants are explicitly modeled (Stegle et al., 2010; Parts et al., 2011; Fusi et al., 2012\nocite{stegle2010bayesian,parts2011joint,fusi2012joint}) have been presented. The basic model has the form:
\begin{equation}
\label{eq:latent_var_regression}
Y=\mathbf{\mu} + SV + XW + \mathbf{\epsilon},
\end{equation}
where $\mathbf{\mu}$ represents phenotype-specific mean terms, $X$ and $S$ denote the observed genotypes and hidden confounders, respectively, with the corresponding regression coefficients collected into the matrices $W$ and $V$. The $\mathbf{\epsilon}$ is a matrix comprising Gaussian i.i.d. noise terms.

The methods differ in the way the model specified in Equation (\ref{eq:latent_var_regression}) is learned. The most thorough way (Fusi et al., 2012\nocite{fusi2012joint}) is to jointly learn the hidden factors and SNPs that influence the phenotype. Here, to facilitate straightforward parallel processing of the genotype blocks, we utilize an approximation implemented in PEER software (Stegle et al., 2010\nocite{stegle2010bayesian}), where the hidden factors are first learned independently of the genotypes and their effects on the phenotypes are removed. The resulting residuals are then used in the place of the phenotypes to test for associations with the genotypes using univariate methods described below.

\subsection{Exhaustive pairwise testing, principal component analysis, and multivariate analysis of variance}

As the baseline we use three methods; the first one is generally used in univariate association analyses, and the other two are straightforward multivariate methods. We use linear regression models to test for assocation between genotype $x_j$ and phenotype $y_i$, $i=1,\ldots,p$ and $j=1,\ldots,q$:
\begin{equation}
y_i = \beta_0 + \beta_1x_j + \epsilon_i.
\end{equation}
The score for an LD-block is taken to be either the smallest p-value for the $\beta_1$ coefficient in any of the genotype-phenotype pairs tested (referred to as \textit{best-pair}), or the average of the corresponding t-test scores over all genotype-phenotype pairs (\textit{avg-pair}). As the second simple baseline method we calculate as many principal components for both genotypes and phenotypes as needed to explain at least 99.5 percent of the variation in each data set. Then we test for associations between the principal components using the univariate test described above. 

As the third baseline method we use the standard multivariate analysis of variance (MANOVA). As with univariate ANOVA, the idea is to divide the total variation in phenotypes to within and between group variation, where the groups are defined by the value of a particular SNP. We apply MANOVA to each SNP in an LD-block in turn, and take the smallest p-value as the score of the block. We obtain the p-value by computing
\begin{equation}
\Lambda=\frac{|W|}{|T|},
\end{equation}
where $T$ (for total) and $W$ (within) are the matrices of sums of squares and products for phenotypes and residuals (after subtracting the group means), respectively. The variable $\Lambda$ has Wilk's Lambda distribution which is approximated using the F-distribution (see, e.g., Mardia, 1979). \nocite{mardia1979multivariate}

\subsection{Metabolomics data set}
\begin{figure}[!tb]
\centering
\includegraphics[trim=1cm 0.5cm 1cm 0.1cm, clip=true, angle=270,width=0.6\textwidth]{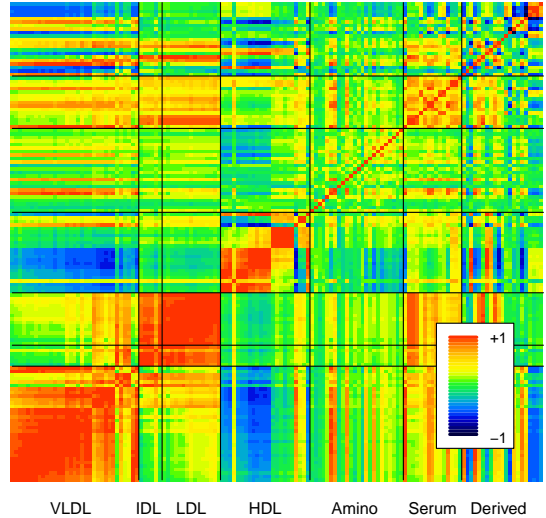}
\caption{\textit{Empirical correlation matrix for the metabolomics data} (before the confounder correction). The subgroups are separated by black lines.}\label{fig:metabol_covariance}
\end{figure}

As the real test case, we used a data set published by Inouye et al. (2010) \nocite{inouye2010metabonomic} which consists of genome-wide SNP data along with metabolomics measurements (for details concerning metabolomics data collection, see Soininen et al., 2009, and Kettunen et al., 2012)\nocite{kettunen2012genome,soininen2009high}. SNPs with low minor allele frequency ($<\hspace{-0.1cm}0.02$) and deviation from Hardy-Weinberg equilibrium ($p\!<\!0.00001$) were removed as a pre-processing step, leaving approximately 550,000 SNPs from 568 unrelated individuals. The metabolomics data set comprised 137 metabolites, most of which represented NMR quantified levels of lipoproteins classified into 4 subclasses (VLDL, IDL, LDL, HDL), together with quantified levels of amino acids, some serum extracts and a set of quantities derived as ratios of the aformentioned metabolites. The final sample size was 509 individuals having both data types. Effects of age, sex, and lipid lowering medication were regressed from the metabolomics data as a pre-processing step (with the latent variable 
regression approach this was done jointly with removing the effects of the hidden confounders). The correlation matrix of the metabolomics data is shown in Figure \ref{fig:metabol_covariance}. A distinguishing characteristic of the data is the blocklike structure composed of groups of highly correlated metabolites.

We defined the genotype block structure by setting block boundaries at loci where adjacent SNPs were more than 0.01 cM apart using genetic map from HapMap Release 22 (NCBI 36) (Frazer et al., 2007)\nocite{frazer2007second}. This resulted in 68,124 LD-blocks with sizes ranging from 1 (32 percent of the blocks comprising 4 percent of the SNPs) to 426 SNPs. Based on graphical inspection, the blocks defined in this way seemed to capture the block-diagonal correlation structure of the genotype data reasonably accurately, although we acknowledge the inherent arbitrariness related to any single fixed cutoff value. By visual comparison, similar divisions could be obtained with the Haploview software (Barrett, 2005\nocite{barrett2005haploview}).

\subsection{Simulated data}
As the basis of our simulations we used two randomly selected LD-blocks from the real genotype data. The genotypes for the simulated data sets were created by sampling with replacement from the set of all available genotypes in these regions. A single SNP was used to generate the effects using a linear model. Afterwards, this causal variant was removed from the genotype data leaving a network of possibly non-linear relationships between the remaining SNPs and phenotypes. This corresponds to the scenario that the true causal variant has not been included in the data set. After generating the effects, the empirical correlation matrix of the metabolomics data was used to simulate correlated additive multivariate Gaussian-distributed noise on top of the simulated phenotypes.

The following factors were varied in the experiments: (1) The size of the LD-block was either 6 or 22 before removing the causal variant. (2) The correlation between the causal variant and the closest SNP in the data set was fixed by manually selecting the causal SNP having the desired correlation with the other SNPs. (3) As the noise correlation matrix we used either the whole metabolomics matrix (137 features) or submatrices corresponding to IDL (6 features) or VLDL (31 features) metabolite subgroups. (4) When the full 137 phenotype features were simulated, the affected traits were selected by mimicking the effects observed in real data (see Figure 1 in Tukiainen et al., 2012\nocite{tukiainen2012detailed}) such that the total number of affected traits was 23. These traits were selected such that they corresponded to three different groups of correlated traits, effects on one of the groups having a different sign from the other two. With smaller numbers of simulated phenotypes, the affected traits were selected analogously 
such that they corresponded to some correlated subclass of the real metabolites. (5) The effect sizes were drawn randomly from the interval $[0.75\beta_{max},\beta_{max}]$, where $\beta_{max}$ is the value reported in Results. Note that actual effects are smaller as the causal variant is not present in the genotype data. Consequently, an upper bound for the proportion of variance explained (PVE) by an effect can be obtained from
\begin{equation}
\text{PVE}=\frac{\beta_{max}^2\text{Var}(x)}{\beta_{max}^2\text{Var}(x)+1},
\end{equation}
where $\text{Var}(x)$ is the variance of the causal variant, which in our simulations was always $<0.5$. Thus, if $\beta_{max}<<1$, the PVE can be roughly bound from above by $0.5\beta_{max}^2$.

\section{Results}

\begin{figure*}[tb]
\centering
\includegraphics[angle=270,width=0.95\textwidth]{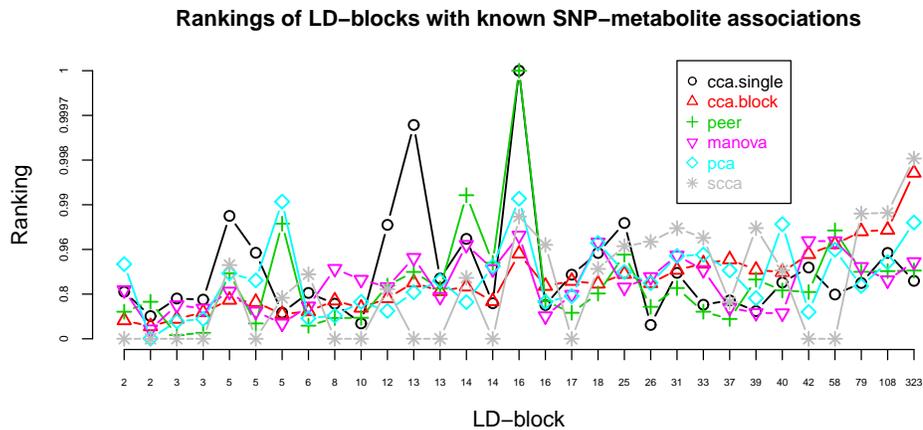}
\caption{\textit{Rankings of LD-blocks with known causal variants}. The rankings of all 68,124 LD-blocks were scaled linearly to the interval $[0,1]$ such that the block with the highest score got rank 1 and the block with the lowest score got rank 0. Rankings of the LD-blocks with known causal variants are shown on the y-axis (notice the logarithmic scale for better separation of values close to unity). The LD-blocks with causal SNPs are ordered according to the size of the block (the number of SNPs in a block shown on the x-axis). Thus, the methods with highest scores in the left end of the x-axis are those with the largest power based on small blocks, whereas those having highest scores in the right end have largest power based on large blocks. To keep the figure readable, results are shown only for the most promising methods for which the median ranking of known causal loci was $\geq0.8$. The ranking of a block was set to zero by sparse CCA (scca) if none of the SNPs in the block got non-zero weights.}\label{fig:block_rankings}
\end{figure*}

\begin{table}[tb]
\centering
\caption{\textit{Comparison of the methods with the real data}. First five columns show summary of the rankings of 31 LD-blocks containing known causal variants. The columns are interpreted as follows: \textit{best}: the number of times the method gave the highest ranking (of all methods considered) to a block with a known causal variant, $\mathit{>0.95}$: the number of blocks with known causal variants ranked among the top 5 percent of all blocks, \textit{median}/\textit{max}/\textit{mean}: the median/maximum/mean rankings of the blocks with known causal variants (after scaling the ranks linearly to $[0,1]$), \textit{sign}: the number of significant findings. The two best methods are shown in bold in each column. The last column, \textit{time}, shows the computation times for a single run (i.e., not including permutation testing). The number of significant findings for GFlasso is not available due to extensive computation time required by the permutation sampling.\vspace{0.4cm}\label{Tab:block_rankings}}
\begin{tabular}{lcccccccc}
\hline
& best & $>0.95$ & median & max & mean & sign & time \\
\hline
\textit{CCA-single} & \bf{8} & \bf{9} & \bf{0.868} & \bf{1.000} & \bf{0.814} & \bf{1} & 7h \\
\textit{CCA-block} & 1 & 6 & 0.864 & 0.998 & \bf{0.806} & 0 & 1h\\
\textit{CCA-block (p-val)} & 1 & 2 & 0.158 & 0.994 & 0.358 & 0 & 1h\\
PEER & 2 & 5 & 0.827 & \bf{1.000} & 0.742 & \bf{1} & 20h\\
\textit{best-pair} & 3 & 8 & 0.777 & \bf{1.000} & 0.743 & 0 & 20h\\
PCA & 3 & 8 & \bf{0.878} & 0.994 & 0.798 & 0 & 11h\\
gfLasso & 1 & 3 & 0.517 & 0.969 & 0.487 & NA & 2,200h\\
MANOVA & 3 & 6 & 0.848 & 0.976 & 0.801 & 0 & 13h\\
Sparse CCA & \bf{9} & \bf{10} & 0.835 & 0.999 & 0.539 & 0 & 8h\\
\hline
\end{tabular}
\end{table}

\subsection{Real data}
We compared the methods using as the ground truth a set of 31 SNPs reported by Tukiainen et al. (2012)\nocite{tukiainen2012detailed}, compactly summarized in Table 13 of Tukiainen (2012)\nocite{tukiainen2012metabolomics}\footnote{During the writing of this article a study using CCA for one SNP at a time was published by Inouye et al. (2012)\nocite{inouye2012novel}. We did not include findings of that article in our baseline, in order not to bias the results in favor of CCA.}. These findings have been obtained using the standard exhaustive pairwise linear regression with a data set of 8,330 individuals (data in our study is a subset of these data).

We first checked whether the methods were able to find the ground truth LD-blocks with significant scores. Multiple-testing corrected thresholds for LD-block scores were obtained by considering maximum scores in 100 repeated analyses with permuted data sets (except for GFlasso due to extensive computation time). With the limited amount of data, only one of the LD-blocks with causal variants scored significantly after the multiple-testing correction. The significant scores were given by methods \textit{CCA-single} and PEER. To further examine how well the different methods are able to highlight the LD-blocks with known causal SNPs, we ordered all LD-blocks using the scores from the methods. The rankings of the blocks with known causal SNPs are shown in Figure \ref{fig:block_rankings} with a summary given in Table \ref{Tab:block_rankings}. Detailed listing of the rankings and the actual scores are given in Supplementary Tables 1 and 2 (including methods not shown here to keep the results uncluttered). The following main conclusions can be drawn from the results:
\begin{enumerate}

\item In general, \textit{CCA-single} and sparse CCA gave higher rankings to the causal blocks than other methods, with \textit{CCA-single} the preferred method in small blocks (consisting of less than fifteen SNPs) and sparse CCA in larger blocks.

\item Sparse CCA applied directly to LD-blocks did not work well. However, when first applied to larger genomic windows, and then computing scores for LD-blocks using the sparsity patterns learned from the window-based analysis, sparse CCA worked very well (see Figure \ref{fig:scca_comparison} in Appendix A for a comparison.) The results shown for sparse CCA are based on this approach.

\item Removing the effects of confounding factors (PEER) before pairwise testing improved the median ranking of the causal blocks. Furthermore, PEER was the only method along with \textit{CCA-single} with which any of the causal blocks got significant scores after the multiple testing correction.

\item However, the top 5 methods (according to the median ranking) were PCA, \textit{CCA-single}, \textit{CCA-block}, MANOVA, and sparse CCA, all of which test for association with all phenotypes jointly. All of these methods were also computationally feasible even with permutation sampling. \footnote{We note that the running times depend on the underlying implementation. For GFlasso we used the executable provided by the authors. Otherwise, straightforward R implementations available either as standard R functions or from the authors of the respective methods were used.}

\item GFlasso did not perform well. We conjecture the reason to be that the Lasso type regularization ceases to work reasonably when the number of SNPs is too small and the SNPs represent a single block with relatively high inter-correlations. Furthermore, the hyperparameters learned by cross-validation were different for different blocks, which may affect the ranking of the blocks by the maximum regression coefficient. Some kind of pooling to learn a single common hyperparameter would seem reasonable and will be worth studying later. Alternatively, we expect that applying regularized regression to larger genomic windows would likely improve the results (as with sparse CCA). However, this approach was not feasible due to extensive computation time.

\end{enumerate}

\begin{figure}[!tb]
\centering
\includegraphics[trim=0cm 0cm 0cm 0cm, clip=true, angle=270,width=0.8\textwidth]{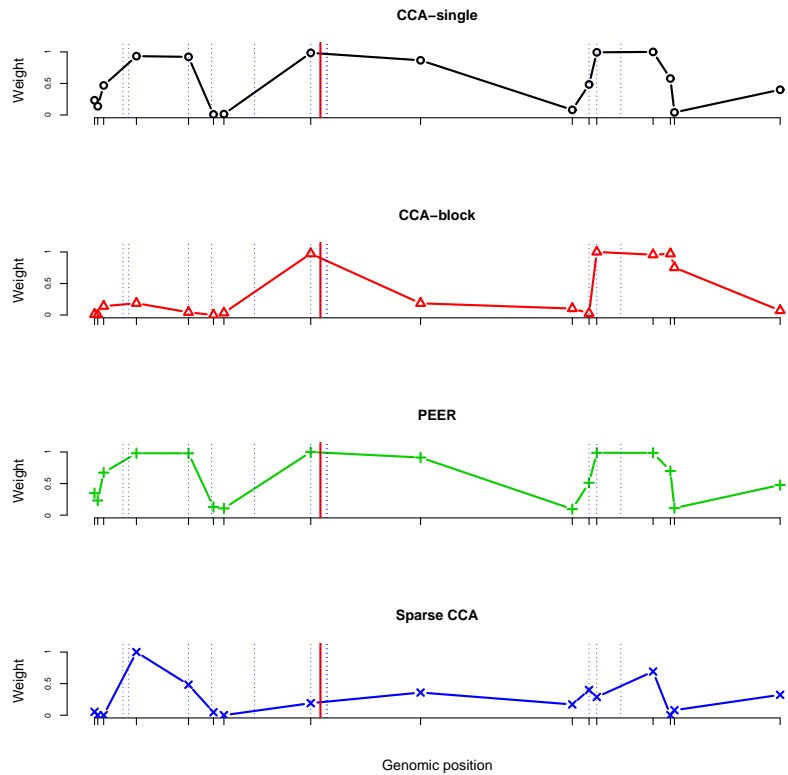}
\caption{\textit{SNP-wise weights for the LD-block significantly associated with the metabolomics profiles}. The location of the known causal lead variant (SNP \textit{rs174547}) is denoted by the red vertical line; however, this SNP was not present in our data set. Other SNPs with reported associations with metabolite traits (from dbSNP, Sherry et al., 2001) are shown with blue vertical lines. With PEER and \textit{CCA-single} the relative weights of SNP importance are obtained by considering the negative logarithm of the corresponding p-values. The absolute values of the canonical weights are shown as the SNP-weights for \textit{CCA-block} and SCCA. All weights are scaled to the interval $[0,1]$.}\label{fig:associated_region}
\end{figure}

To illustrate the results with the real data on a more detailed level, Figure \ref{fig:associated_region} shows SNP-wise results from \textit{CCA-single} and PEER for the LD-block that obtained significant scores by these methods. The SNP-wise weights from \textit{CCA-block} and sparse CCA are added for comparison. Because we used permutation sampling to obtain multiple-testing corrected significance scores for the whole block, the SNP-wise scores are not as such comparable between methods (to avoid confusion, scores from each method are therefore scaled to $[0,1]$ interval in the figure). However, it is of interest to investigate the relevance of different SNPs in a block to see how well the location of the causal SNP is identified. The main conclusion from the figure is that sparse CCA and \textit{CCA-block} give high scores to fewer SNPs in the block. This is not surprising as the methods that score each SNP separately (\textit{CCA-single} and PEER) are expected to give high scores to all SNPs that are correlated with the causal SNP, whereas \textit{CCA-block} and especially sparse CCA are expected to pick a sparse combination of SNPs that is maximally correlated with the phenotypes. This general trend was verified by plotting the proportion of high-scoring SNPs in each block, see Figure \ref{fig:proportions} in Appendix A.

Using the p-values from Rao's approximation to rank the LD-blocks (\textit{CCA-block (p-val)} in Table \ref{Tab:block_rankings}) seems to work badly compared to using directly the maximum canonical correlation (\textit{CCA-block}). Indeed, when we investigated the p-values more closely, many of them were very close to unity, indicating that the distribution of the test scores does not match with the assumptions underlying the significance test. Tang and Ferreira (2012) used genotype pruning to reduce collinearity between SNPs to resolve this problem. However, since in this type of data both the genotypes and the phenotypes are highly correlated, the pruning strategy seems too conservative to be motivated from the biological perspective. The downside of using the maximum canonical correlation is that larger blocks are \textit{a priori} more likely to obtain high canonical correlations, leading to reduced power with smaller blocks. Introducing sparsity among the SNPs with sparse CCA seems to improve results in this respect, although \textit{CCA-single}, which forms the score of a whole block using the single most correlated SNP only, is still preferred with small blocks.

Finally, Supplementary Table 1 shows SNP-wise rankings relative to all 550,000 SNPs for 8 causal SNPs (as opposed to blocks) that were included in our data set. Although these rankings are not directly comparable to the block rankings, it is notable that the SNP-wise CCA utilized by \nocite{inouye2012novel}Inouye et al. (2012) gives for 7/8 SNPs lower rankings than the blockwise CCA formulations for the corresponding blocks.

\nocite{sherry2001dbsnp}

\subsection{Simulations}
We investigated the power of the methods to detect associations in two different simulation setups. Throughout, we used data sets simulated with effect size set to zero to determine significance thresholds yielding the empirical false positive rate equal to 0.05.

\subsubsection{Whole metabolomics profile simulation}

\begin{figure}[tb]
\centering
\includegraphics[trim=0cm 0.5cm 0cm 0.1cm, clip=true, angle=270,width=0.92\textwidth]{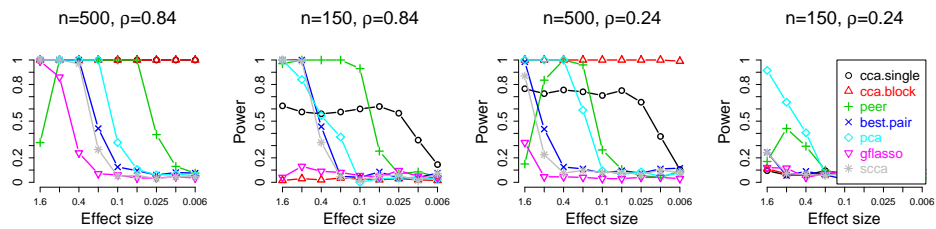}
\caption{\textit{Power of different methods in the whole metabolomics profile simulation scenario.} Factors that were varied include $n$, the number of individuals, $\rho$, the largest correlation between the (not included) causal variant and any of the SNPs included in the data set, and the effect size. The value of the effect size shown in the figure is relative to one standard deviation of noise. Corresponding to each parameter configuration, 200 data sets with the specified effect and 200 data sets with zero effect were analyzed. The empirical false positive rate was fixed to 0.05 by selecting an appropriate significance threshold using the data sets simulated under no effect. In the panel on the left the curve for \textit{CCA-single} is behind \textit{CCA-block}.}\label{fig:simu_scenario1}
\end{figure}

In the first setup, we investigated how the size of the effect, the number samples, and the correlation between the causal variant and the closest observed SNP affect the power of the different methods to detect associations. We fixed the dimension of the phenotype vector to 137 (using the whole metabolomics correlation matrix to generate the noise). The results are summarized in Figure \ref{fig:simu_scenario1}. The following conclusions can be drawn:
\begin{enumerate}

\item The classical CCA is the best method when of number of samples is large enough relative to the dimension of the genotype and phenotype blocks tested, but breaks down otherwise. The difference between the \textit{CCA-single} and \textit{CCA-block} is intuitive: if none of the SNPs present is highly correlated with the causal variant, the \textit{CCA-block} is capable of better utilizing the information in the whole genotype block, outperforming \textit{CCA-single}. On the other hand, when some observed SNP is highly correlated with a (single) causal variant, \textit{CCA-block} has no advantage over \textit{CCA-single}. We also note that, in line with what was observed with the real data, the sparse CCA does not work well when testing for association of a single, relatively small LD-block (here 22 SNPs) against the phenotypes.

\item PEER is among top-three methods in all setups and seems to be the method of choice when the number of observations is small compared to the dimension, that is, the realm for which it was originally developed. The power of PEER decreases as the effect size gets large enough. This behavior is expected and can be explained by the fact that PEER starts explaining away the true effect with latent confounders. A possible solution suggested by Stegle et al. (2010)\nocite{stegle2010bayesian} would be to iterate between learning the effects and latent confounders. Alternatively, joint learning of the effects and confounders (Fusi et al., 2012\nocite{fusi2012joint}) would likely improve the results in this respect.

\item GFlasso did not work well in our simulation setup, the reasons being the same as discussed for the real data. Discouraged by these results, we did not include GFlasso in the second simulation setup (see below) to save some computation time. To check that the conclusion is not specific to GFlasso, but more generally to Lasso-type regression, we additionally ran the whole metabolomics simulation scenario with another Lasso regression method, the Sparse Group Lasso (SGL) by Hastie and Tibshirani (2012); however, we note that this method is not specifically designed for multivariate responses. Therfore, we ran SGL separately for each phenotype and took the maximum over the these as the overall score. The results are shown in Figure \ref{fig:additional} (Appendix A) and provide support to the stated conclusion.

\item PCA performs better than the other methods when no single SNP with high correlation with the causal variant is present in the data set, and sample size is small relative to the dimension of data (the fourth panel in Figure \ref{fig:simu_scenario1}). However, in this setup, even PCA seems to require unrealistically large effects sizes before satisfactory behaviour can be expected.

\end{enumerate}

\begin{figure}[!tb]
\centering
\includegraphics[trim=0cm 0.5cm 0cm 0.1cm, clip=true, angle=270, angle=0, width=0.77\textwidth]{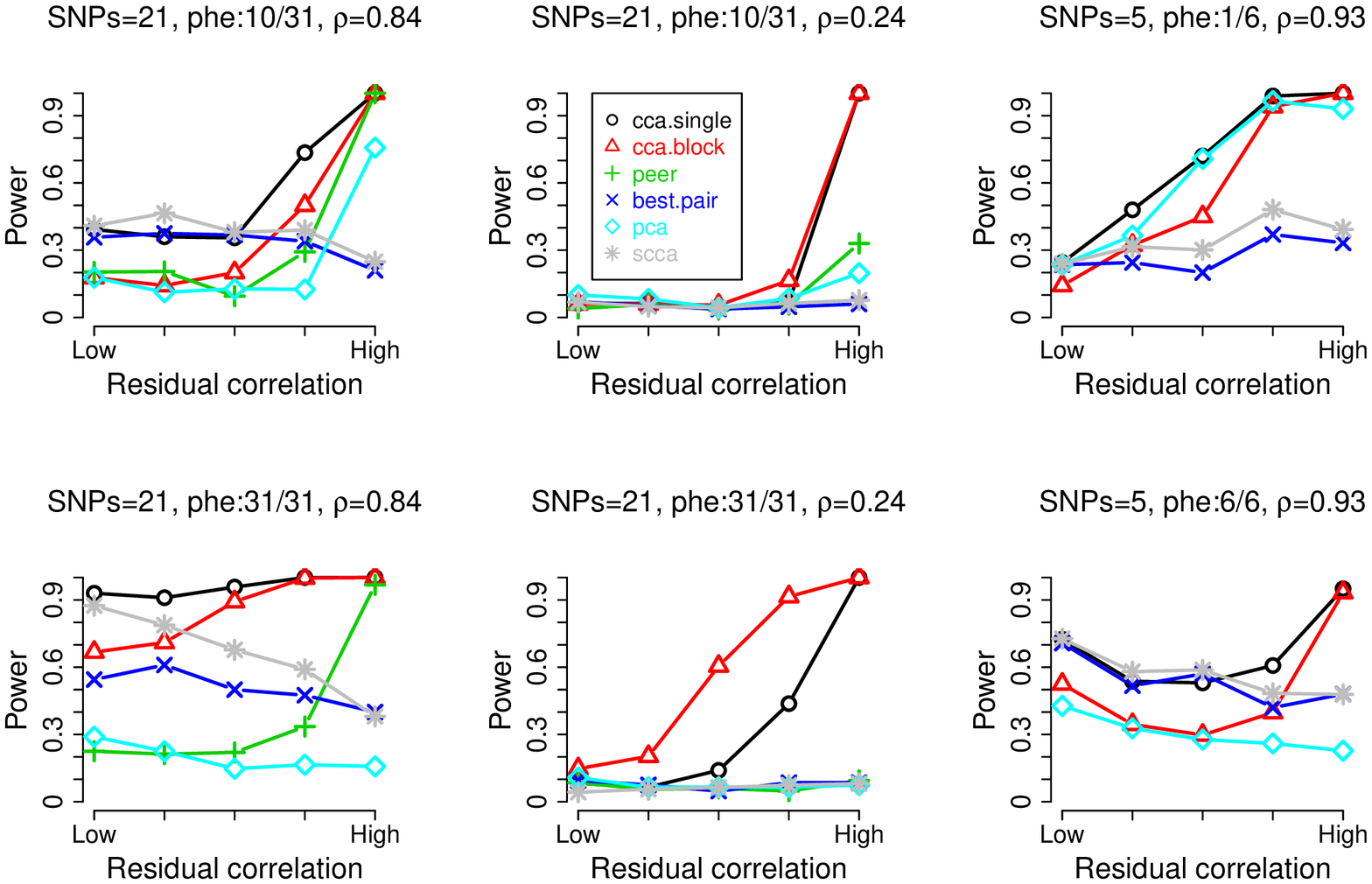}
\caption{\textit{Power of different methods in the metabolite subgroup simulation scenario.} The titles of the panels show the number of \textit{SNPs}, and the maximum correlation between the causal variant and the most correlated SNP present in data, $\mathit{\rho}$. Text \textit{phe:x/y} in the title tells the number of phenotypic traits (\textit{y}) and the number of traits affected by the causal variant (\textit{x}). High residual correlation corresponds to using the empirical correlation matrix of a specific group of highly correlated metabolites to simulate the noise. The less correlated data sets are created by raising the correlation matrix to powers 10, 20, 40 and 80, effectively pulling the off-diagonal elements of the correlation matrix towards zero. Corresponding to each parameter configuration, 400 data sets with the specified effect and 400 data sets with zero effect were analyzed. The empirical false positive rate was fixed to 0.05 using the data sets simulated under no effect. PEER converged in less 
than 5 percent of the data sets with 6 phenotypes (the third column), and is not shown for these data sets. }\label{fig:simu_scenario2}
\end{figure}

Figures \ref{fig:pval_comparison}, \ref{fig:additional}, and \ref{fig:two_causal} in Appendix A provide additional insight to the behavior of the methods. Figure \ref{fig:pval_comparison} compares the power of \textit{CCA-block} using the canonical correlation directly vs. p-values calculated using either the Bartlett's or Rao's approximation as the test score. The results are in line with what was observed with the real data, namely that using the canonical correlation is preferred to using the p-values.

Figure \ref{fig:additional} shows results similar to Figure \ref{fig:simu_scenario1} for methods not included in Figure \ref{fig:simu_scenario1} for clarity. The most interesting of these is MANOVA, which shows very good performance. A closer inspection reveals that the performance profile of MANOVA is very similar to \textit{CCA-single}, except that the power is always lower than or equal to \textit{CCA-single}. The same behaviour was also observed in all other simulations (exact results not shown). The similarity of MANOVA and \textit{CCA-single} can be expected because, although the objective functions are not exactly the same, both are essentially trying to find the best SNP in a block to explain variation in all phenotypes. MANOVA does this by trying to explain the variance with group means, groups defined by individuals having different SNP values (0,1, or 2). \textit{CCA-single}, on the other hand, tries to find a linear relationship between the SNP and the phenotypes, resulting in a more restricted model with increased power.

Finally, we checked how sensitive the results are with respect to the assumption that only one causal SNP is used to simulate the effects (and subsequently removed from the data set before the analysis). Figure \ref{fig:two_causal} shows results when two causal SNPs were used, and neither was removed from data. The general trends are very similar to those observed in Figure \ref{fig:simu_scenario1}, with the main difference being the improved performance of \textit{CCA-single} with the lower sampler size. This is not surprising as the scenario with the true causal SNPs present in data is favourable to \textit{CCA-single} which tries to find the best single SNP to explain the variation in the phenotypes.

\subsubsection{Metabolite subgroup simulation}

In the second setup, we investigated the effect of residual (i.e. noise) correlation, number of features in the genotype and phenotype data sets, and the number of affected traits on the power to detect associations. As the basis of simulating the phenotypes we used the empirical correlation matrix of IDL (6 features) or VLDL (31 features) metabolite subgroups. The results from this setup are summarized in Figure \ref{fig:simu_scenario2}.

\begin{figure}[tb]
\begin{centering}
\includegraphics[trim=0cm 0cm 0cm 0cm, clip=true,width=0.8\textwidth]{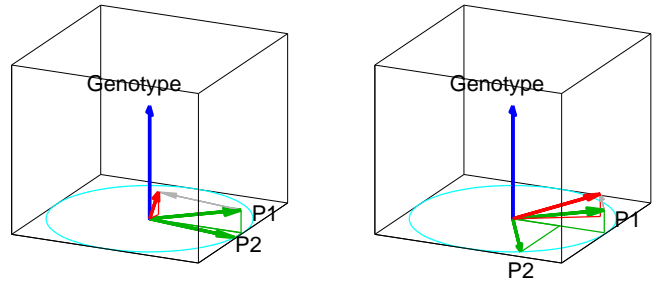}
\caption{\textit{Graphical illustration for why CCA works better with correlated data}. The green vectors $P1$ and $P2$ represent two phenotypes in $n$-dimensional space. The blue vector represents the genotype. Note that $P1$ has a component that points to the same direction as the genotype. The red arrow shows the linear combination of the phenotypes that has the minimum angle with the genotype. In the left panel $P1$ and $P2$ are highly correlated, i.e. pointing roughly to the same direction, in the right panel, the correlation between $P1$ and $P2$ is small.}\label{fig:cca_illustration}
\end{centering}
\end{figure}

The most obvious trend is the improved performance of CCA when the residual correlation increases. Some intuition to this behavior can be obtained by interpreting the $n$ observations for variables $y_i$ and $x_j$ as vectors in an $n$-dimensional space, and noticing that the correlation between any two variables equals the cosine of the angle between the correponding vectors (see, e.g., Mardia, 1979\nocite{mardia1979multivariate}). In particular, the canonical correlation defined in Equation (\ref{eq:can_cor}) is the cosine of the angle between linear combinations $Xa$ and $Yb$. Thus, canonical correlation analysis attempts to minimize the angle between the linear combinations. Figure \ref{fig:cca_illustration} illustrates this in a simplified situation with two phenotypes and a single genotype. One of the phenotypes, $P1$, is correlated with the genotype, that is, it has a component that points to the same direction as the genotype. The other phenotype, $P2$, falls on the plane perpendicular to the genotype, and is 
therefore uncorrelated with the genotype. When $P1$ and $P2$ are correlated (left panel), $P2$ can be used to cancel the component of $P1$ that is perpendicular to the genotype, leading to higher canonical correlation. In the other extreme $P1$ and $P2$ would be completely uncorrelated (perpendicular to each other), in which case $P2$ would be of no use in making the angle between $P1$ and the genotype smaller. Thus, better use can be made of the component pointing to the same direction with the genotype if the phenotypes are correlated, which explains the higher canonical correlations with correlated phenotypes.

Similar reasoning can be used to explain the behavior noticed by others (Ferreira and Purcell, 2009; Tang and Ferreira, 2012; Waaijenborg et al., 2008\nocite{ferreira2009multivariate,tang2012gene,waaijenborg2008quantifying}) that the power of CCA decreases if the genotype affects simultaneously all phenotypes. If all phenotypes are positively correlated and have an equal component to the direction of the genotype, it is easy to visualize that that component is removed simultaneously when subtracting the component perpendicular to the genotype. In our simulations this behavior was not prominent. The explanation is that, although in our simulations the causal SNP affected a group of correlated SNPs, the exact effect sizes were selected randomly from a specific interval, thus not canceling each other completely when taking linear combinations.

\section{Discussion}
In this work, we have investigated methods that can be used to detect small effects in GWASs with high-dimensional phenotypes, by taking the whole phenotype vector jointly into account. Our main conclusion, supported by both simulations and analysis of a real data set, is that canonical correlation analysis appears to be the most powerful approach for this purpose. If the number of samples is reduced to the level of the dimension of the genotype or phenotype group to be tested, regression models with latent confounders (such as implemented in PEER) also seem promising. Furthermore, with the real data, PEER and CCA were the only methods in our study with which any of the known causal variants got significant scores after multiple-testing correction.

We allowed the methods to combine information in the genotype data in a computationally feasible way, by dividing the genome into blocks of correlated SNPs. Compared to analysing the whole genome jointly, the possibility to process the blocks in parallel makes the methods considered in this study computationally feasible even with permutation sampling to obtain multiple-testing corrected significance thresholds. Even if CCA analysis using jointly the full set of genotypes was possible, the interpretation of the canonical components might be tedious, as discussed e.g. by Waaijenborg (2008)\nocite{waaijenborg2008quantifying}. Considering a block of neighboring SNPs at a time focuses the putative effect on certain part of the genome making the interpretation easier. Further, picking the SNP with the largest coefficient in the canonical correlation vector seems a promising way of recovering the SNPs most correlated with the phenotypes, the strategy also suggested by Naylor et al. (2010)\nocite{naylor2010using}. On the other hand, compared to analysing each SNP separately, the block-wise approach reduces the dimensionality of the problem, lessening the multiple-testing problems. In our simulations, we saw increased power to detect causal variants that were not included in the data set when the whole block of genotypes was tested jointly. Further, with the real data the blocks with causal variants were ranked higher by CCA (relative to all blocks) than the actual causal SNPs (relative to all SNPs).

Although the approach where all SNPs in a block are jointly tested against all phenotypes with CCA had the highest power in the simulations, it has the problem that the canonical correlation, which was used as the test score, depends on the size of the block and overfits when the number of samples is close to the dimensionality of data. A methodological solution to this problem is to regularize the model to obtain a sparse solution giving non-zero weights to only a subset of SNPs and/or phenotypes. We included in our analysis two variants of this concept: one where a sparse combination of SNPs was learned using formal regularization techniques, the other where the score of the block was simply taken to be the maximum canonical correlation between any single SNP and the phenotypes. These two approaches were also the best methods on the real data, the former ranking large and the latter small LD-blocks on average higher than other methods. However, it is worth pointing out that for satisfactory behavior the regularized CCA needed first to be applied to larger windows and not to the LD-blocks directly. After this, the learned sparsity structure in the SNP data was employed heuristically to calculate the scores for the LD-blocks using the classical CCA. It is notable that although sliding window approaches are relatively often suggested for analysing large data sets, there does not seem to exist a commonly used formal strategy for comparing and combining results between different windows with differing hyperparameters. Developing a rigorous way to do this would be highly beneficial when analysing large data sets in practice, both with sparse CCA as well as with other sparse methods, such as regression models.

The regularized CCA utilized in this article is not the only alternative to extend the usability of CCA to high-dimensional data sets under small $n$ large $p$ conditions. Other approaches include Bayesian (Wang, 2007; Klami and Kaski, 2007; Virtanen et al., 2011\nocite{wang2007variational,klami2007local,virtanen2011bayesian}) and kernel CCA (Hardoon et al., 2004\nocite{hardoon2004canonical}). These types of methods often rely on computationally extensive techniques, such as cross-validation, to learn the hyperparameters of the model. Investigation of possible gains at the price of increased computational burden with realistic-sized GWAS data sets remains an open question. Besides CCA, there are other methods that have been proposed for detecting associations between multivariate genetic data sets, including sparse partial least squares (PLS, L{\'e} Cao et al., 2008) and sparse reduced rank regression (RRR, Vounou et al., 2010). Whereas CCA maximizes the correlation between two data sets, PLS maximizes the covariance. However, as our phenotype data includes measurements on different scales, it is reasonable to scale the data before the analysis, after which we expect the difference between CCA and PLS to be small. RRR, on the other hand, tries to find low-dimensional projections of the genotype data to be used as regressors to predict the phenotypes. For a discussion on the conceptual differences between CCA, PLS and RRR, see Vounou et al. (2010). \nocite{lecao2008sparse,vounou2010discovering}

Throughout the paper, we have advocated the strategy of dividing the genotype side into computationally manageable parts. Indeed, we feel that developing multivariate methods that attempt to take the whole genotype into account simultaneously are destined to fail, as present-day GWAS data sets may contain tens of millions of SNPs, and even loading such data sets in computer memory as a whole may be impossible. On the other hand, we have demonstrated that by dividing the problem appropriately into parts, even computationally intensive techniques, such as cross-validation to learn the hyperparameters of sparse CCA, may be feasible. In addition to the growing dimensionality of the genotype data, the data can grow both in terms of the number of phenotypes as well as the number of individuals present in a data set. Here we have considered a phenotype consisting of 135 traits. As long as the number of individuals is large relative to the number of phenotypes, the classical CCA and MANOVA are feasible. Besides PCA and exhaustive pairwise testing, which work well with very high-dimensional data, also sparse CCA and PEER have been demonstrated with thousands of phenotypes; however, usually with a relatively small number, say a few hundred, of individuals.

The largest GWASs may nowadays comprise hundreds of thousands of individuals. For example, Deloukas et al. (2012) analysed 63,756 cases and 130,681 controls for coronary artery disease. Besides dividing the genotypes into manageable parts, a strategy commonly used in international GWAS projects with data from several collaborators is to divide the individuals into subsets to be analyzed separately. The results are then combined over the data sets using meta-analysis techniques, see, e.g., Thompson et al. (2012). The simplest way to do this is to estimate the effect and its variance separately for each data set and to pool the results by taking the precision (inverse variance) weighted average of the individual effect estimates. Assumption of a common fixed effect size underlies this procedure, and can be relaxed by applying random effects models, for example. We were not able to find studies were outcome of a multivariate statistical analysis using CCA would have been combined over several data sets in a similar fashion. At the very least, this would require that estimates of a test score (canonical correlation) with approximated errors could be formed for each data set. Especially with sparse CCA analytical error bounds for the canonical correlation are not available, and even with the classical CCA their existence depends on assumptions such as normality of the data. On the other hand, sampling based techniques, such as bootstrap, might be used to approximate the variances of the canonical correlation estimates. These could then be used to compute the pooled estimate as the weighted average of individual estimates; however, such an approach clearly requires further investigation.\nocite{deloukas2012large,thompson2011meta}

\section*{Supplementary Data}
The Supplementary Data referred to in the text is available as a single zip-file from \\ \url{http://users.ics.aalto.fi/pemartti/high_dimensional_supplementary/}. The file contains Supplementary Tables 1-2 and the captions to the supplementary Tables.

\section*{Acknowledgements}
The authors thank the National Institute for Health and Welfare for providing the real data and Boreal Plant Breeding Ltd. for insightful discussions. This work was financially supported by the Academy of Finland (grant number 251170 to the Finnish Centre of Excellence in Computational Inference Research COIN; grant number 259272 to P.M.; grant number 118065 to DILGOM study), the European Research Council (grant number 239784 to J.C), and the Research Foundation of Helsinki University of Technology (grant to J.G.).

\newpage

\section*{Appendix A: Additional figures}

\begin{figure}[h!]
\begin{centering}
\includegraphics[trim=0cm 0cm 0cm 0cm, clip=true, angle=270, width=0.85\textwidth]{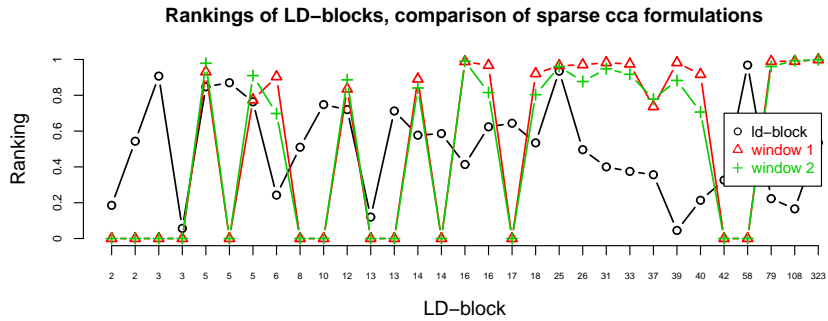}
\caption{\textit{Comparison of three different sparse CCA formulations with the real data}. The figure shows rankings of the LD-blocks with known causal SNPs, and is interpreted exactly as Figure \ref{fig:block_rankings}. The formulations are: \textit{ld-block}: the score of a block is obtained by computing the canonical correlation between the SNPs in the block and the phenotypes using the sparse CCA method. \textit{window 1}: The score of a block is obtained by computing as the first step the sparse canonical correlation between a large window with several LD-blocks (containing more than 2,000 SNPs in total) and all phenotypes, and then, as the second step, computing for each block the classical canonical correlation between all SNPs that obtained non-zero weights in the block and all phenotypes. \textit{window 2}: The score is obtained as in \textit{window 1} except that in the second step only the phenotypes which had non-zero weights in the first step are used to compute the classical canonical correlation.}\label{fig:scca_comparison}
\end{centering}
\end{figure}

\begin{figure}[h!]
\begin{centering}
\includegraphics[trim=0cm 0cm 0cm 0cm, clip=true, angle=270, width=0.85\textwidth]{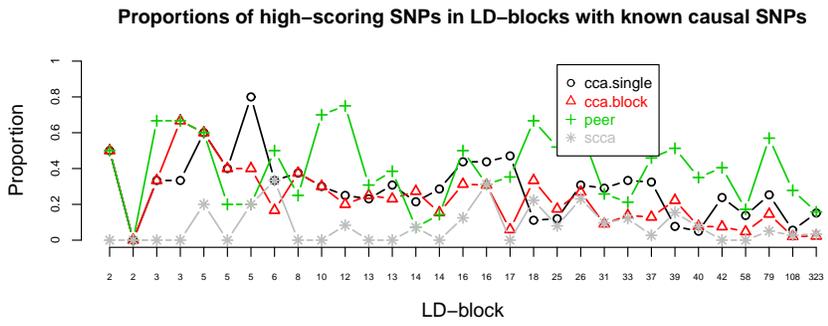}
\caption{\textit{Proportions of high-scoring SNPs in the LD-blocks with known causal SNPs}. A high-scoring SNP is defined as a SNP whose score is larger than the average of the smallest and the largest scores in a block. The scores of the different methods are defined as in Figure \ref{fig:associated_region}.}\label{fig:proportions}
\end{centering}
\end{figure}

\begin{figure}[h!]
\begin{centering}
\includegraphics[trim=0cm 0cm 0cm 0cm, clip=true, angle=270, width=0.92\textwidth]{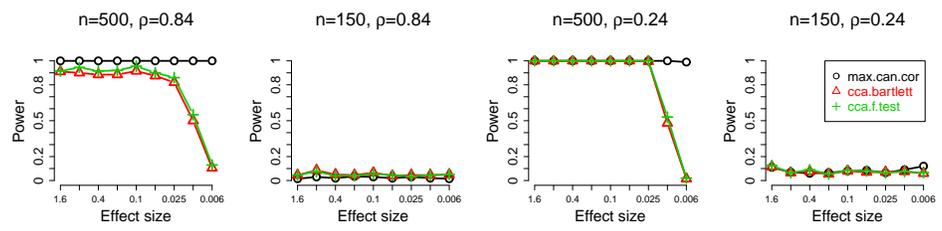}
\caption{\textit{A comparison of three possible scoring alternatives for the CCA-block method using the whole metabolomics profile simulation setup}. The figure is interpreted exactly as Figure \ref{fig:simu_scenario1}, except that the list of methods that are compared is different. The following methods are included: \textit{cca.block}: perform CCA between the genotype and phenotype blocks, and take the maximum canonical correlation as the test score. \textit{cca.bartlett}: use (-logarithm of) p-value obtained for all canonical correlations using Bartlett's approximation. \textit{cca.f.test}: is the same as \textit{cca.bartlett}, except that Rao's F-approximation is used to compute the p-value.}\label{fig:pval_comparison}
\end{centering}
\end{figure}

\begin{figure}[h!]
\begin{centering}
\includegraphics[trim=0cm 0cm 0cm 0cm, clip=true, angle=270, width=0.92\textwidth]{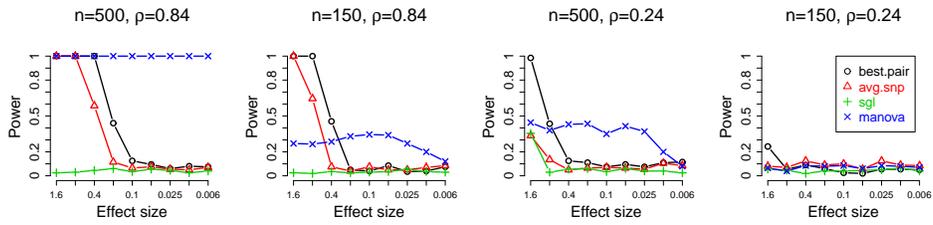}
\caption{\textit{Power of additional methods in the whole metabolomics profile simulation scenario.} The figure is interpreted similarly to Figure \ref{fig:simu_scenario1}, only different methods are shown.}\label{fig:additional}
\end{centering}
\end{figure}

\begin{figure}[h!]
\begin{centering}
\includegraphics[trim=0cm 0cm 0cm 0cm, clip=true, angle=270, width=0.6\textwidth]{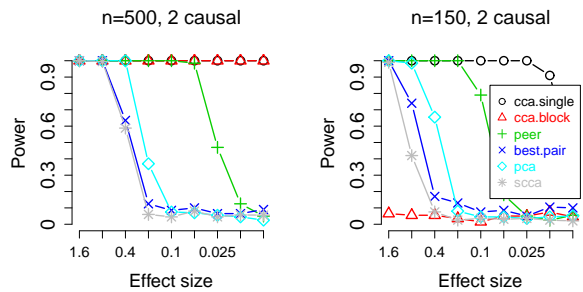}
\caption{\textit{Power in the simulation scenario with two SNPs}. The figure is interpreted similarly to Figure \ref{fig:simu_scenario1}, except that two causal SNPs were used when simulating the effects and, unlike in Figure \ref{fig:simu_scenario1}, the causal SNPs were not removed from the genotype data set.}\label{fig:two_causal}
\end{centering}
\end{figure}

\clearpage
\bibliographystyle{DeGruyter}
\bibliography{cca_bibliography.bib}

\end{document}